\documentclass[conference]{IEEEtran}
\IEEEoverridecommandlockouts
\usepackage{cite}
\usepackage{amsmath,amssymb,amsfonts}
\usepackage{algorithmic,algorithm}
\usepackage{graphicx}
\usepackage{textcomp}
\usepackage{xcolor}
\usepackage{tabulary}
\def\BibTeX{{\rm B\kern-.05em{\sc i\kern-.025em b}\kern-.08em
    T\kern-.1667em\lower.7ex\hbox{E}\kern-.125emX}}
    
\usepackage{mathtools} 
\usepackage{xspace}

\newcommand{\herm}{^\text{H}}

\newcommand{\trans}{^\text{T}}

\newcommand{\bs}{\mathbf{s}}

\newcommand{\bH}{\mathbf{H}}

\newcommand{\bR}{\mathbf{R}}

\newcommand{\ba}{\mathbf{a}}

\newcommand{\bv}{\mathbf{v}}

\newcommand{\bu}{\mathbf{u}}

\newcommand{\CN}{\mathcal{CN}}

\DeclarePairedDelimiter\floor{\lfloor}{\rfloor}

\newcommand{\boundellipse}[3]
{(#1) ellipse [x radius=#2,y radius=#3]
}

\newcommand{\EX}[1]{\mathsf{E}\left\{{#1}\right\}}

\newcommand{\tauc}{\tau_\mathrm{c}}

\newcommand{\LoS}{^\textsf{LoS}}
\newcommand{\AP}{_\textsf{AP}}
\newcommand{\UE}{_\textsf{UE}}
\newcommand{\s}{_\textsf{s}}

\newcommand{\ds}{\displaystyle}
\newcommand{\wt}{\widetilde}

\newcommand{\comment}[1]{}
\newcommand{\imgj}{\mathsf{j}} 
\newcommand{\eul}{\mathsf{e}}
\newcommand{\taus}{\tau_\mathsf{s}}

\begin{document}

\title{User-centric Handover in mmWave Cell-Free Massive MIMO with User Mobility}
\author{
\IEEEauthorblockN{Carmen D'Andrea, Giovanni Interdonato and Stefano Buzzi}
\IEEEauthorblockA{Dipartimento di Ingegneria Elettrica e dell'Informazione, University of Cassino and Southern Latium, Cassino, Italy \\
\{carmen.dandrea, giovanni.interdonato, buzzi\}@unicas.it}\thanks{The authors are also with the Consorzio Nazionale Interuniversitario
per le Telecomunicazioni (CNIT), 43124, Parma, Italy. This paper has been supported by the Italian Ministry of Education University and Research (MIUR) Project ``Dipartimenti di Eccellenza 2018-2022'' and by the MIUR PRIN 2017 Project ``LiquidEdge''.}}

\maketitle

\begin{abstract}
The coupling between cell-free massive multiple-input multiple-output (MIMO) systems operating at millimeter-wave (mmWave) carrier frequencies and user mobility is considered in this paper. First of all, a mmWave channel is introduced taking into account the user mobility and the impact of the channel aging. Then, three beamforming techniques are proposed in the considered scenario, along with a dynamic user association technique (handover): starting from a user-centric association between each mobile device and a cluster of access points (APs), a rule for updating the APs cluster is formulated and analyzed. Numerical results reveal that the proposed beamforming and user association techniques are effective in the considered scenario.
\end{abstract}


\section{Introduction}

Cell-free massive MIMO is deemed a key technology for \textit{beyond}-5G systems~\cite{Zhang2020,Rajatheva2020,Matthaiou2020} mainly for its great ability to provide a uniformly excellent spectral efficiency (SE) throughout the network  and ubiquitous coverage~\cite{Ngo2017b,Interdonato2019}. 
Decentralizing operations such as channel estimation, power control and, possibly, precoding/combining prevents to create any bottleneck in the system, while confining the signal co-processing within a handful of properly, dynamically selected APs, implementing thereby a user-centric system~\cite{Buzzi2019c}, makes cell-free massive MIMO practical. Erasing the cell boundaries when transmitting/receiving data requires, however, smooth coordination and accurate synchronization among the APs.

The SE improvements that cell-free massive MIMO can provide over co-located massive MIMO and small cells are significant in the sub-6 GHz frequency bands, especially in terms of service fairness. In the mmWave bands, where achieving high spectral efficiency is not a concern thanks to the large available bandwidth, the distributed cell-free operation can certainly better cope with the hostile propagation environment at such high frequencies compared to a co-located system, as the presence of many serving APs in the user equipment (UE)'s proximity enhances coverage and link reliability. 
A few recent works have investigated the potential marriage between cell-free massive MIMO and mmWave under various viewpoints: energy efficiency maximization~\cite{Alonzo2019}, analysis in case of capacity-constrained fronthaul links~\cite{Femenias2019}, energy-efficient strategies to cope with a non-uniform spatial traffic distribution~\cite{Morales2020}. 

An aspect that deserves a closer look in this context is the impact on the performance of the UE mobility. As the UE moves, the propagation channels towards the APs are subject to time variations, and thereby the available channel state information (CSI) at the AP ages with time, effect called \textit{channel aging}. The channel aging effects have been extensively analyzed for co-located massive MIMO~\cite{Heath_ChannelAging2013,Papazafeiropoulos_TVT2017} (and references therein), and recently for cell-free massive MIMO~\cite{ZhengJ2020} operating at sub-6 GHz frequency bands. While for mmWave bands, the effects of the channel aging have been evaluated for \textit{Hetnet} MIMO systems in~\cite{Papazafeiropoulos_WCNC2019}. In most of the works in the literature, the channel aging is considered within a transmission block (or resource block) and usually intended as the mismatch between the acquired CSI prior the transmission, used for beamforming/combining, and the actual CSI at the transmission/reception time (channel estimation error aside).         

\textbf{Contributions:} The novelty of this work consists in studying the coupling between cell-free massive MIMO, mmWave and user mobility. First of all, we introduce the mmWave channel model taking into account the user mobility and the impact of the channel aging. Then, we describe three beamforming techniques assuming different levels of complexity at the transmitter and receiver and propose a dynamic user association technique, which starts from a traditional user-centric approach and updates the serving cluster of APs for each user in order to reduce the number of unnecessary handovers. Numerical results reveal that the proposed beamforming and user association techniques are effective in the considered scenario.  

\section{System model}
Let us consider a cell-free massive MIMO system operating at mmWave where $M$ multi-antenna APs simultaneously serve $K$ multi-antenna UE in the same time-frequency resources in time-division duplex (TDD) mode. We assume that both the APs and the UE are equipped with a uniform linear array (ULA) with random orientations, and steering angles taking on values in $[-\pi/2,\pi/2]$. Each AP and each UE has $N\AP$ and $N\UE$ antenna elements, respectively, with $MN\AP > KN\UE$. Finally, let $n\s < \min(N\AP,N\UE)$ be the number of streams multiplexed over the MIMO channel.

Unlike most of the works in the literature, we consider a channel that (almost) continuously evolves over time according to the variation of the UE locations with respect to the APs, hence according to a model that takes into account the UE mobility.

\subsection{Channel Model}

\subsubsection{mmWave MIMO Channel Representation}

\comment{We assume that the adopted modulation format is the orthogonal frequency division multiplexing (OFDM). Let us denote by $t_0$ the duration of an OFDM symbol. 
We denote by $B$ the overall available bandwidth, and by $\Delta f$ the subcarrier spacing for the OFDM signal. 
This implies that the number of subcarriers is $N_C=B/\Delta f$. The duration of the OFDM symbol is taken equal to $1/\Delta f + \tau_{CP}$, with $\tau_{CP}$ the length of the cyclic prefix,  chosen not smaller than the largest multipath delay spread in the system. The time line is divided into four different units, namely length of the OFDM symbol $t_0$, length of the time slot containing 14 OFDM symbols $T_{\rm slot}$, channel coherence time $T_{C}$ and slow-fading coherence time $T_S$. 
The channel estimation is performed every $T_{C}$ and the user association, that is essentially based on the slow-fading variation, is performed every $T_S$.} 

The mmWave MIMO narrow-band channel between UE $k$ and AP $m$ consists in a dominant line-of-sight (LoS) path and $L_{km} \ll \min(N\AP,N\UE)$ non-LoS (NLoS) paths due to the presence of scattering clusters. At the time instant $n$, $n=\{1,2,3\ldots\}$, the channel is modelled by the complex-valued $(N\AP\!\times\!N\UE)$-dimensional matrix
\begin{equation}
\bH_{mk}[n]=\! \varsigma\!\!\sum_{\ell=1}^{L_{mk}}\alpha_{mk}^{(\ell)}[n] \ba\AP(\phi_{mk}^{(\ell)}[n])\ba\UE\herm(\theta_{mk}^{(\ell)}[n])\!+\! \bH\LoS_{mk}[n],
\label{eq:channelmodel_km} 
\end{equation} 
where $\varsigma = \ds \sqrt{N\AP N\UE}$ is a normalization factor, 
$\alpha^{(\ell)}_{mk}[n]\sim\CN(0,\beta^{(\ell)}_{mk}[n])$ is the complex-valued gain of the $\ell$-th path with strength (which reflects the path loss) denoted by $\beta^{(\ell)}_{mk}[n]$. 
The unit-norm ULA steering vectors at UE $k$ and AP $m$ are denoted by $\ba\UE(\theta_{mk}^{(\ell)}[n])$ and $\ba\AP(\phi_{mk}^{(\ell)}[n])$, respectively, highlighting the dependency on the angle of departure (AoD), $\theta_{mk}^{(\ell)}[n]$, and the angle of arrival (AoA), $\phi_{mk}^{(\ell)}[n]$, of the $\ell$-th path. Assuming half-wavelength spacing between the antenna elements of the ULA, the steering vectors for a generic AoD and AoA are given by
\begin{align}
\ba\AP(\phi) &= \frac{1}{\sqrt{N\AP}}[1, \; \eul^{\imgj \pi \sin \phi}, \; \ldots, \; \eul^{\imgj \pi (N\AP-1)\sin \phi}]\trans, \label{eq:a_AP} \\
\ba\UE(\theta) &= \frac{1}{\sqrt{N\UE}} [1, \; \eul^{\imgj \pi \sin \theta}, \; \ldots, \; \eul^{\imgj \pi (N\UE-1)\sin \theta}]\trans.
\label{eq:a_UE}
\end{align}
The LoS component of the channel matrix at the $n$th time instant, $\bH\LoS_{mk}[n]$, is given by
\begin{equation} \label{eq:LoS:channel_matrix}
\bH\LoS_{mk}[n] \! = \! \varsigma \epsilon(d\LoS_{mk}[n]) \varrho\LoS_{mk}[n] \ba\AP(\phi\LoS_{mk}[n])\ba\UE\herm(\theta\LoS_{mk}[n]),  
\end{equation}
where $$\varrho\LoS_{mk}[n] \triangleq \sqrt{\beta\LoS_{mk}[n]} \eul^{\imgj \vartheta_{mk}[n]},$$ with $\vartheta_{mk}[n]\!\sim\!\mathcal{U}(0,2\pi)$, and $\epsilon(d\LoS_{mk}[n])\!\in\!\{0,1\}$ denotes the binary blockage variable characterizing the LoS link between AP $m$ and UE $k$ with length $d\LoS_{mk}[n]$, and indicating whether the LoS path, at the $n$th time instant, is obstructed. The LoS channel strength is denoted by $\beta\LoS_{mk}[n] \gg \beta^{(\ell)}_{mk}[n], \ell = 1, \ldots, L_{mk}$.

\comment{
We assume a number of total scatterers, $N_{\rm s}$ say, common to all the APs and users and uniformly distributed in the system. In order to model the blockage, we assume that the communication between the $m$-th AP and the $k$-th user takes place via the $n$-th scatterer, i.e., the $n$-th scatterer is one of the effective $L_{k,m}$ contributing in the channel in Eq. \eqref{eq:channelmodel_km}, if the rays between the $m$-th AP and the $n$-th scatterer and the $k$-th user and the $n$-th scatterer \emph{simultaneously} exist. We assume a link exists between two entities, in our case one AP/UE and one scatterer, if they are in LoS, with a probability, $P_{\rm LOS}(d)$ depending on the distance between the two entities, $d$ say.

In the vast majority of channel measurements campaigns in millimeter wave bands the number of rays between two entities is very low (insert references), i.e., the millimeter wave channel is sparse. In this paper, we focus in the case in which only one predominant line of sight (LoS) path is assumed between one user and one AP.}

\subsubsection{User Mobility}
UE mobility leads to temporal variations in the propagation environment resulting in a channel that ages with time. This channel aging is much more significant in mmWave systems than in sub-6 GHz systems as a time variation results in a larger phase variation at high frequency bands than at low frequency bands.

In this work, we assume a block-fading channel model wherein the path gains---the fast fading component of the channel---stay constant within a resource block of length $\tauc$ channel uses (or time instants) and vary over the resource blocks. Whereas the ULA steering vectors and the channel variances stay constant over multiple resource blocks, for an interval with duration $\taus>\tauc$ time instants.
The temporal evolution of the NLoS path gains over the resource blocks is modeled through a first order autoregressive model \cite{Heath_ChannelAging2013,Papazafeiropoulos_TVT2017}, such that a path gain at the current time instant $\alpha^{(\ell)}_{mk}[n]$, with $n=\{1,2,3\ldots\}$, is a function of its previous state $\alpha^{(\ell)}_{mk}[n\!-\!1]$ and an innovation component $\wt{\alpha}^{(\ell)}_{mk}[n]$ as
\begin{align} \label{eq:aging:alpha}
\begin{cases}
\delta_{mk}[n] \alpha^{(\ell)}_{mk}[n\!-\!1]\!+\! \bar{\delta}_{mk}[n] \wt{\alpha}^{(\ell)}_{mk}[n], &\text{if } n\!\!\!\!\mod\tauc \!=\! 1, \\
\alpha^{(\ell)}_{mk}[\floor{n/\tauc}], &\text{otherwise,}
\end{cases}
\end{align}
with $\wt{\alpha}^{(\ell)}_{mk}[n] \sim \CN(0,\beta_{mk}^{(\ell)}[n])$ and independent of $\alpha_{mk}^{(\ell)}[n-1]$, $\floor{\cdot}$ is the \textit{floor} function,
$\ds \bar{\delta}_{mk}[n] = \sqrt{1-\delta_{mk}^2[n]}$, and $\delta_{mk}[n]$ represents the temporal correlation coefficient of UE $k$ with respect to AP $m$, which follows the Jake's model
\begin{equation}
\delta_{mk}[n]= J_0 \left( 2 \pi f_{\mathsf{D},mk} n T \right), 
\end{equation}
with $J_0(\cdot)$ being the zeroth-order Bessel function of the first kind, and $T$ being the sampling period. $f_{\mathsf{D},mk}$ is the maximum Doppler frequency between the $k$-th UE and the $m$-th AP given by
$
f_{\mathsf{D},mk} =\nu_{k,m} f_\mathsf{c}/c \, ,
$
where $\nu_{mk}$ is the radial velocity of UE $k$ with respect to AP $m$ in meters per second (m/s), $c = 3 \cdot 10^8$ m/s is the speed of light, and $f_\mathsf{c}$ is the carrier frequency in Hz.

Importantly, the temporal variation of the channel variances, both for LoS and NLoS, is modeled as
\begin{align} \label{eq:aging:beta}
  \beta^{(\ell)}_{mk}[n]=
  \begin{cases}
  	f_{\mathsf{PL}}(d_{mk}^{(\ell)}[n]), & \text{if } n \!\!\!\! \mod \taus = 1, \\
  	\beta^{(\ell)}_{mk}[\floor{n/\taus}], & \text{otherwise},
  \end{cases}
\end{align}
namely the path loss is assumed to stay constant within $\taus$ channel uses.  The function $f_{\mathsf{PL}}(d_{mk}^{(\ell)}[n])$ denotes the path loss function which is characterized by the distance between the $k$-th UE and the $m$-th AP over the $\ell$-th path at the $n$-th time instant, $d_{k,m}^{(\ell)}[n]$.

Similarly, the AoAs and the AoDs (both for LoS and NLoS) vary every $\taus$ channel uses, and their temporal evolution is modeled as in Eq.~\eqref{eq:aging:angles}, shown at the top of the next page, where $f_{\mathsf{G}}(\cdot)$ represents the geometric function depending on the propagation scenario. Whereas the phase shift of the LoS channel component, $\vartheta_{mk}[n]$, takes on a new random value uniformly distributed on the interval $[0, 2\pi]$ every resource block but stays constant within $\tau_c$ channel uses.
\begin{figure*}[!t]
  \begin{align}   \label{eq:aging:angles}
  \left(\phi_{mk}^{(\ell)}[n], \theta_{mk}^{(\ell)}[n]\right) =  
  \begin{cases}
  f_{\mathsf{G}}(d_{mk}^{(\ell)}[n]), &\text{if } n \!\!\!\! \mod \taus = 1 \\
  \left(\phi_{mk}^{(\ell)}[\floor{n/\taus}], \theta_{mk}^{(\ell)}[\floor{n/\taus}]\right), &\text{otherwise}.
  \end{cases}
  \end{align}
\hrulefill
\end{figure*}   
    
The variation of the $k$-th UE's location from the time instant $n$ to $n+1$ is modelled as 
\begin{align}
x_k[n+1] &= x_k[n]+ v_{k}T_\mathsf{s} \cos\varphi_{k}, \\
y_k[n+1] &= x_k[n]+ v_{k}T_\mathsf{s} \sin\varphi_{k},
\end{align}
where $\varphi_{k} \in [0, 2 \pi]$ is the direction of the movement and $v_{k}$ is the speed of the $k$-th UE with respect to a reference system centred on the UE location and assuming that $\varphi_{k}$ and $v_{k}$ remain constant over $\taus$ channel uses. A new realization of the UE's location determines a new set of distances $\{d_{mk}^{(\ell)}\}$ and, as a consequence, a new set of coefficients $\{\beta^{(\ell)}_{mk}\}$.

\subsection{Downlink Signal Model} \label{DL_model}

Without loss of of generality, we assume $n_\mathsf{s} = 1$, i.e., only one stream is transmitted for each AP-UE pair. Let $\bu_{mk}[n]$ be the $N\AP$-dimensional precoding vector between AP $m$ and the UE $k$, and let $\bv_{k}[n]$ be the $N\UE$-dimensional combining vector used at the $k$-th user. We also define $a_{mk}[n]$ as a binary parameter to indicate whether the $m$-th AP serves the $k$-th UE. More specifically,
\begin{align}
a_{mk}[n]=
\begin{cases}
1 \quad &\text{if AP } m \text{ serves UE } k,\\
0 \quad &\text{otherwise}.
\end{cases}
\end{align}
The downlink data signal sent by the $m$-th AP is given by the following $N_{AP}$-dimensional vector-valued waveform
\begin{equation}
\bs_{m}[n]=\ds \sum\nolimits_{j=1}^{K} a_{mj}[n] \sqrt{\eta_{mj}[n]}\bu_{mj}[n]x_{mj}[n],
\end{equation}
where $x_{mj}[n]$ is the data symbol intended for the $j$-th UE, $\EX{|x_{mj}[n]|^2} = 1, \forall j, m$, and $\eta_{mj}[n]$ is the transmit power. 
The signal received at the $k$-th UE after the combining operation can be then expressed as
\begin{align}
\label{eq:received_k_n}
r_k[n] \! &= \! \sum\nolimits_{m=1}^M a_{mk}[n] \sqrt{\eta_{mk}[n]} \mathbf{v}\herm_{k}[n]  \mathbf{H}\herm_{mk}[n] \mathbf{u}_{mk}[n] x_{mk}[n] \nonumber \\
&\quad +\!\!\!\sum _{m=1}^M \sum_{j \neq k}^K a_{mj}[n] \sqrt{\eta_{mj}[n]} \bv\herm_{k}[n] \bH\herm_{mk}[n]\bu_{mj}[n]x_{mj}[n] \nonumber \\ 
& \quad + z_{k}[n],
\end{align}
where $z_{k}[n]\sim \CN(0,\sigma^2_k[n])$ is additive noise.

\subsubsection*{Precoding and Combining}
We analyze three beamforming techniques: long-term beamforming (LTB), short-term beamforming (STB) and analog beamforming (ABF). We consider two different implementations for both LTB and STB: $(i)$ digital implementation (DI) and $(ii)$ constant-modulus implementation (CI), which is also suitable for a hybrid (analog/digital) implementation. In the case of LTB, each AP and UE performs long-term precoding and long-term combining, respectively~\cite{Akdeniz_JSAC2014}, which requires only the knowledge of the channel variances, AoAs and AoDs of the channel. 
We thereby define the $\left(N\AP \times N\AP \right)$-dimensional covariance matrix 
\begin{align}
\mathbf{R}_{mk}^{\mathsf{(AP)}}[n] &= \EX{\bH_{mk}[n]\bH\herm_{mk}[n]} \label{eq:LTBF-AP} \\ 
& = \varsigma^2 \epsilon(d\LoS_{mk}[n]) \beta\LoS_{mk}[n] \ba\AP(\phi\LoS_{mk}[n])\ba\AP\herm(\phi\LoS_{mk}[n]) \nonumber \\
&\quad + \varsigma^2 \sum\limits_{\ell=1}^{L_{mk}}\beta_{mk}^{(\ell)}[n] \ba\AP(\phi_{mk}^{(\ell)}[n]) \ba\AP\herm(\phi_{mk}^{(\ell)}[n]),
\label{eq:R_AP}
\end{align}
and the $(N\UE \times N\UE)$-dimensional covariance matrix 
\begin{align}
\mathbf{R}_{mk}^{\mathsf{(UE)}}[n]&= \EX{\bH\herm_{mk}[n] \bH_{mk}[n]} \label{eq:LTBF-UE} \\ 
& = \varsigma^2 \epsilon(d\LoS_{mk}[n]) \beta\LoS_{mk}[n] \ba\UE(\theta\LoS_{mk}[n])\ba\UE\herm(\theta\LoS_{mk}[n]) \nonumber \\
&\quad + \varsigma^2 \sum_{\ell=1}^{L_{mk}}\beta_{mk}^{(\ell)}[n] \ba\UE(\theta_{mk}^{(\ell)}[n]) \ba\UE\herm(\theta_{mk}^{(\ell)}[n]).
\label{eq:R_UE}
\end{align}
In the case of LTB-DI, the precoding vector $\mathbf{u}_{mk}^{\mathsf{(DI)}}[n]$ is equal to the predominant eigenvector of the matrix $\mathbf{R}_{mk}^{\mathsf{(AP)}}[n]$, while the combining vector $\mathbf{v}_{k}^{\mathsf{(DI)}}[n]$ is equal to the predominant eigenvector of the matrix
\begin{equation}
\label{eq:R_UEk}
\overline{\bR}_k^{\mathsf{(UE)}}[n] = \sum\nolimits_{m=1}^M a_{mk}[n] \bR_{mk}^{\mathsf{(UE)}}[n].
\end{equation}
In the LTB-CI, the generic entry of the beamforming and combining vectors has constant modulus, and phase as that of its digital counterpart, i.e.,
\begin{equation}
\begin{array}{llll}
\mathbf{u}_{mk}^{\mathsf{(CI)}}[n]=\ds \frac{1}{\sqrt{N\AP}}e^{\imgj\angle{\mathbf{u}_{mk}^{\mathsf{(DI)}}[n]}}, \quad
\mathbf{v}_{k}^{\mathsf{(CI)}}[n]=\ds \frac{1}{\sqrt{N\UE}}e^{\imgj\angle{\mathbf{v}^{\mathsf{(DI)}}_{k}[n]}}.
\end{array}
\label{eq:CI}
\end{equation}   
where $\angle{\cdot}$ denotes the phase of the argument.

STB can be performed if and only if we ideally assume that both APs and UEs have knowledge of the instantaneous realization of the channel $\bH_{mk}[n]$. 
In the case of STB-DI, the precoding vector is given by the predominant eigenvector of the matrix $\bH_{mk}[n]\bH\herm_{mk}[n]$, whereas the combining vector at the UE is given by the predominant eigenvector of the matrix defined as in~\eqref{eq:R_UEk} but with no expectation in~\eqref{eq:LTBF-UE}.
The expression of the STB-CI does not differ from that of LTB-CI, the precoding and combining vectors follow Eq. \eqref{eq:CI} with $\mathbf{u}_{mk}^{\mathsf{(DI)}}[n]$ and $\mathbf{v}^{\mathsf{(DI)}}_{k}[n]$ evaluated according to the STB-DI.
\comment{We thereby define the $\left(N\AP \times N\AP \right)$-dimensional matrix 
\begin{equation}
\label{eq:R_AP_STB}
\widetilde{\mathbf{R}}_{mk}^{\mathsf{(AP)}}[n]= \bH_{mk}[n]\bH\herm_{mk}[n] 
\end{equation}
and the $(N\UE \times N\UE)$-dimensional matrix 
\begin{equation}
\label{eq:R_UE_STB}
\widetilde{\mathbf{R}}_{mk}^{\mathsf{(UE)}}[n]= \bH\herm_{mk}[n] \bH_{mk}[n]
\end{equation}
The precoding vector for the DI, $\mathbf{u}_{mk}^{\mathsf{(DI)}}[n]$ say, is equal to the predominant eigenvector of the matrix $\widetilde{\mathbf{R}}_{mk}^{\mathsf{(AP)}}[n]$, while the combining vector, $\mathbf{v}_{k}^{\mathsf{(DI)}}[n]$ say, is equal to the predominant eigenvector of the matrix
\begin{equation}
\widetilde{\bar{\mathbf{R}}}_k^{\mathsf{(UE)}}[n] = \sum_{m=1}^M a_{mk}[n] \widetilde{\mathbf{R}}_{mk}^{\mathsf{(UE)}}[n].
\end{equation}}
Finally, the precoding and combining vectors for the ABF scheme are set as the steering vectors in~\eqref{eq:a_AP} and~\eqref{eq:a_UE} corresponding to the AoD and AoA of the strongest path, respectively, for each AP-UE pair.

\section{User association technique}
In order to detail the user association technique, we firstly define the \textit{channel strength indicator} for the communication between the $m$-th AP and the $k$-th UE, denoted by $\rho_{mk}[n]$, which is equal to the predominant eigenvalue of the channel covariance matrix. 
Any UE is served by a \textit{user-centric cluster} of APs, consisting of the $N_{\mathsf{UC}}$ APs with the largest channel strength indicator. We let $O^{(n)}_{k} \, : \, \{1,\ldots, M \} \rightarrow \{1,\ldots, M \}$ denote the operator sorting the AP indices by the descending order of the vector $[\rho_{1k}[n], \ldots, \rho_{Mk}[n]]$, such that $O^{(n)}_{k}(1)$ gives the index of the AP with largest $\rho_{mk}[n]$.  
The set $\mathcal{M}_k[n]$ of the $N_{\mathsf{UC}}$ APs serving the $k$-th user is then given by
\begin{equation}
\mathcal{M}_k[n]=\{ O^{(n)}_{k}(1), O^{(n)}_{k}(2), \ldots , O^{(n)}_{k}(N_{\mathsf{UC}}) \}.
\end{equation}
Hence, if $m \in \mathcal{M}_k[n]$, then $a_{mk}[n]=1$, and $a_{mk}[n]=0$ otherwise. The evolution of the UE's location due to the mobility changes the relative channel strength indicators between APs and UEs and thus the UE-to-APs associations. In this paper, we propose the following UE-to-APs association finalized to reduce the number of instantaneous handovers which can negatively affect the system performance.

As we have assumed that the UE's mobility produces a non-negligible effect on the channel every $\taus$ samples, the UE-to-APs association updates accordingly, that is
\begin{equation}
\overline{\rho}_{mk}^{(q)}=\rho_{mk}\left[n: q=\ds \floor{n/\taus} \right].
\end{equation}
In order to reduce the number of instantaneous handovers, we define two \textit{hysteresis parameters}: a value,  $\zeta_{\mathsf{HO}}$ say, and a control number, $N_{\mathsf{HO}}$ say, used to dynamically manage the user association.
The proposed dynamic user association is summarized in Algorithm \ref{User_association_Dynamic}.

\begin{algorithm}
	
	\caption{The proposed dynamic user association}
	
	\begin{algorithmic}[1]
		
		\label{User_association_Dynamic}
		\STATE  Start with $q=0$ and perform the initial user-centric approach based on the channel strength indicators $\overline{\rho}_{mk}^{(0)}$ and obtain the set $\mathcal{M}_k[0]$.
		
		\FOR{$q=1,2,3,\ldots$}
		
		\STATE Evaluate the new $\overline{\rho}_{mk}^{(q)}, \, \forall m,k$;

	 	\FOR{$k=1,\ldots,K$}

		 \STATE Define $$m^-_k(q-1)= \arg \min_{m \in \mathcal{M}_k[q-1] }\overline{\rho}_{mk}^{(q-1)}$$ and $$m^+_k(q)=\arg \max_{m \notin \mathcal{M}_k[q-1] }\overline{\rho}_{mk}^{(q)}.$$
		
		 \STATE Replace the worst AP in $\mathcal{M}_k[q+N_{\mathsf{HO}}]$ \emph{only if} 
		\begin{equation}
			\ds\frac{\overline{\rho}_{m^+_k(h+1)k}^{(h+1)}-\overline{\rho}_{m^-_k(h)k}^{(h)}}{\overline{\rho}_{m^-_k(h)k}^{(h)}} > \zeta_{\mathsf{HO}},
			\label{eq:hysteresis_condition}
		\end{equation}
		with $h=q-1, q, q+1,\ldots,q+ N_{\mathsf{HO}}-1$.
		
		\ENDFOR
		
	\ENDFOR
	
	\end{algorithmic}
	
\end{algorithm}

The proposed dynamic user-association procedure is aimed at controlling and reducing the unnecessary handovers. Unnecessary handovers can be a consequence of temporary variations of the large scale fading coefficient caused by the user mobility. In order to reduce them, we consider the hysteresis parameters and change the set of the serving cluster of a generic user only if the condition in Eq. \eqref{eq:hysteresis_condition} is satisfied. The hysteresis condition in Eq. \eqref{eq:hysteresis_condition} controls, for a generic user in the system, that the AP that we are removing from the serving cluster is not temporarily in a blockage condition but the user is significantly moving away from it.

\section{Numerical Results}
In our simulation setup, we assume an orthogonal frequency-division multiplexing (OFDM) system operating at $f_0=28$ GHz with bandwidth $W = 500$ MHz. The subcarrier spacing is 480 kHz, and assuming that the length of the cyclic prefix is 7\% of the OFDM symbol duration, i.e., $\tau_{\mathsf{CP}}\Delta_f=0.07$, we obtain $t_0=2.23~\mu$s and $N_C=1024$ subcarriers. The antenna heights are $10$ m at the APs and $1.65$ m at the UEs as in \cite{Buzzi2019c}. The additive thermal noise has a power spectral density of $-174$ dBm/Hz, while the front-end receiver at the APs and at the UEs has a noise figure of $9$ dB. In order to avoid to exceed the simulation area due to mobility, the APs are uniformly distributed in a square area of 850 m $\times$ 850 m, while the initial position of the UEs are uniformly distributed in an inner centred square area of 350 m $\times$ 350 m. 
We set $K=20$, $M=120$, $N\AP=32$ and $N\UE=16$. We assume a number of total scatterers, $N_{\mathsf{s}}=200$, common to all the APs and UEs and uniformly distributed in the simulation area where the APs are located. To model the blockage, we assume that the communication between the $m$-th AP and the $k$-th UE takes place via the $n$-th scatterer, i.e., the $n$-th scatterer is one of the effective $L_{mk}$ contributing to the channel in Eq. \eqref{eq:channelmodel_km}, if the rays between the $m$-th AP and the $n$-th scatterer and the $k$-th UE and the $n$-th scatterer \emph{simultaneously} exist. We assume there exists a link between two entities, in our case one AP/UE and one scatterer, if they are in LoS, with a probability $P_{\mathsf{LoS}}(d)$ being function of the distance $d$ between the two entities, and defined as \cite{ghosh20165g_WP,5G3GPP}
\begin{equation}
P_{\mathsf{LoS}}(d)=\text{min}\left(20/d,1\right)\left(1-e^{-d/39}\right)+e^{-d/39} \; .
\end{equation} 
The function $f_{\mathsf{PL}}\left(d_{mk}^{(\ell)}[n]\right)$ follows the Urban Microcellular (UMi) Street-Canyon model in \cite{ghosh20165g_WP}.
Note that, given the locations of scatterers, UEs and APs in the system, the corresponding AoAs, AoDs, path lengths and $f_{\mathsf{G}}(z)$ in Eq. \eqref{eq:aging:angles} are obtained with straightforward geometric relationships.
In the downlink data transmission phase, we assume equal power allocation, i.e., letting $P_m^{\mathsf{dl}}$ be the maximum transmit power at the $m$-th AP, we set $\eta_{mk}[n]=P_m^{\mathsf{dl}}/\sum\nolimits_{j=1}^K a_{mj}$.

We evaluate the system performance by assuming that both the APs and the UEs perfectly know the channel variances, the AoAs and the AoDs whenever required.

\begin{figure}[!t]
	\centering
	\includegraphics[scale=0.6]{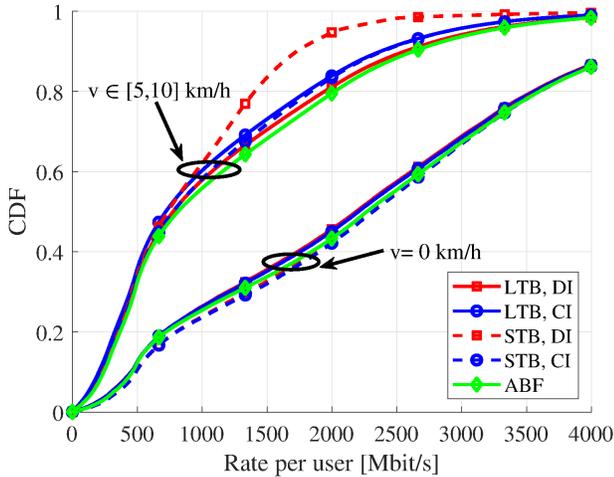}
	\caption{CDF of downlink rate per user, comparison between the considered beamforming techniques in the cases of $v=0$ km/h and $v \in [5,10]$ km/h. Parameters: $N_{\mathsf{UC}}=5$, $\zeta_{\mathsf{HO}}=5$ \% and $N_{\mathsf{HO}}=10$. }
	\label{Fig:Rate_vmax10}
\end{figure}

The signal-to-interference-plus-noise ratio (SINR) at the $n$-th time instant, in case of perfect channel state information knowledge, is denoted by $\text{SINR}_k[n]$ and from Eq. \eqref{eq:received_k_n} it is equal to
\begin{align} \label{eq:SINR_k}
\frac{\left|\sum\limits_{m=1}^M a_{mk}^{[n]} \sqrt{\eta_{mk}[n]} \bv_k[n]\herm \bH\herm_{mk}[n]\bu_{mk}[n] \right|^2}{\sum\limits_{j \neq k}^K \left|\sum\limits_{m=1}^M  a_{mj}[n] \sqrt{\eta_{mj}[n]} \bv_k[n]\herm \bH\herm_{mk}[n] \bu_{mj}[n]\right|^2 + \sigma^2_k[n]}.
\end{align}
Hence, a downlink achievable rate for the UE $k$ is given by
$
\text{R}_k[n]= W \log_2\left( 1+ \text{SINR}_k[n] \right).
$

In Fig. \ref{Fig:Rate_vmax10} we report the CDF of the downlink rate per user in the considered scenario using the beamforming techniques detailed in Section \ref{DL_model}, both for the case of no-moving users, i.e., $v=0$ km/h, and slow-moving users with $v \in [5,10]$ km/h. For the proposed dynamic user association we assume that the hysteresis parameters are $\zeta_{\mathsf{HO}}=5$ \% and $N_{\mathsf{HO}}=10$ and the dimension of the serving cluster for each user is $N_{\mathsf{UC}}=5$. Inspecting the results we can firstly note that the user mobility degrades the performance in terms of rate per user with respect to the case of no-moving users. Regarding the beamforming techniques, we can see that the LTB, which is more suitable for a practical implementation than the STB, offers good performance. We can also see that the ABF, which focuses the power only in the main AoA and AoD for each pair UE-AP is effective, especially in the case of user mobility. This behaviour is due to the sparse nature of the mmWave channel which allows us, on one hand, to focus the transmit power in few directions and, on the other hand, to reduce the interference in the system. Similar insights can be also obtained by the results in Table \ref{table:v_max_50} which reports the median Rate (MR) and 95\%-likely rate (95-R) of the system assuming the different beamforming techniques in the cases of slow-moving users with $v \in [5,10]$ km/h and fast-moving users with $v \in [20,50]$ km/h.

\begin{table}[!t]
\centering                                                                                 
\renewcommand{\arraystretch}{1.3}
\setlength{\tabcolsep}{5.5pt}
\caption{Median Rate (MR) and 95\%-likely Rate (95-R) in Mbps}
\begin{tabular}{|c|c|c|c|c|c|c|c|c|}
\hline
UE speed                     & \multicolumn{4}{c|}{$v \in [5,10] $ km/h}                 & \multicolumn{4}{c|}{$v \in [20,50]$ km/h}                \\ \hline
Rate                     & \multicolumn{2}{c|}{MR} & \multicolumn{2}{c|}{95-R} & \multicolumn{2}{c|}{MR} & \multicolumn{2}{c|}{95-R} \\ \hline
$\zeta_{\mathsf{HO}}$                              & 5\%          & 20\%         & 5\%           & 20\%          & 5\%          & 20\%         & 5\%            & 20\%         \\ \hline
LTB, DI                       & 777        & 791        & 135         & 136         & 659        & 682        & 109          & 112        \\ \hline
LTB, CI                       & 720        & 734        & 122         & 123         & 617        & 638        & 102          & 105        \\ \hline
\multicolumn{1}{|l|}{STB, DI} & 736        & 749        & 149         & 151         & 648        & 669        & 122          & 126        \\ \hline
\multicolumn{1}{|l|}{STB, CI} & 776        & 791        & 146         & 148         & 649        & 684        & 123          & 127        \\ \hline
\multicolumn{1}{|l|}{ABF}      & 814        & 830        & 156         & 158         & 684        & 709        & 129          & 133        \\ \hline
\end{tabular}
	\label{table:v_max_50}                            
\end{table} 

\section{Conclusions and future work}
In this paper we considered a cell-free massive MIMO scenario at mmWave with UE mobility. First of all, we introduced the channel model taking into account the user mobility and the impact of the channel aging and described three beamforming techniques assuming different levels of complexity. Taking into account the impact of the user mobility, we proposed a dynamic user association technique, which starts from a traditional user-centric approach and updates the serving cluster of APs for each user in order to reduce the number of unnecessary handovers. Numerical results reveal that the proposed techniques are effective and motivate us to deepen on this topic. In particular, the presented results assumed perfect knowledge of the channel variances, the AoAs and the AoDs and thus, further investigations are required on the impact of channel estimation error and the use of a protocol which exploits the sparse and autoregressive nature of the channel realizations.

\bibliography{Cell_free_references}
\bibliographystyle{ieeetran}
\end{document}